\documentclass[%
 reprint,
showpacs,
 amsmath,amssymb,
 aps,
 pra,
longbibliography,
floatfix
]{revtex4-2}

\usepackage[utf8]{inputenc}	%Tillader danske tegn
\usepackage[T1]{fontenc}	%Tillader danske tegn
\usepackage[]{graphicx}		%Tillader indsættelse af billeder
\usepackage{xcolor}
\usepackage{gensymb}
\usepackage{braket}
\usepackage{comment}
\usepackage{microtype}
\usepackage[english]{babel}
\usepackage{physics}

\renewcommand{\b}{\hat{b}}

\renewcommand{\a}{\hat{a}}

\renewcommand{\L}{\hat{L}}

\renewcommand{\H}{\hat{H}}

\begin{document}

\title{Jaynes-Cummings interaction with a traveling light pulse}

\author{Victor Rueskov Christiansen}
\email{victorrc@phys.au.dk}
\author{Mads Middelhede Lund}
\email{mml@phys.au.dk}
\affiliation{
Center for Complex Quantum Systems, Department of Physics and Astronomy,
Aarhus University, Ny Munkegade 120, DK-8000 Aarhus C, Denmark}

\author{Fan Yang}
\email{fan.yang@nbi.ku.dk}\author{Klaus M{\o}lmer}
\email{klaus.molmer@nbi.ku.dk}
\affiliation{Center for Hybrid Quantum Networks, Niels Bohr Institute, University of Copenhagen, Blegdamsvej 17, DK-2100 Copenhagen, Denmark}

\date{\today}

\bigskip

\begin{abstract}
The Jaynes-Cummings model provides a simple and accurate description of the interaction between a two-level quantum emitter and a single mode of quantum radiation. Due to the multimode continuum of eigenmodes in free space and in waveguides, the Jaynes-Cummings model should not be expected to properly describe the interaction between an emitter and a traveling pulse of quantum radiation. In this article, we review a cascaded quantum system approach that accurately describes the interaction of a quantum system with an incident quantum pulse of radiation. This approach leads to different formulations of the theory, each of a similar structure as the Jaynes-Cummings model but with important modifications.
\end{abstract}

\maketitle
\noindent

\section{Introduction}

In their seminal article, \emph{Comparison of Quantum and Semiclassical Theories with Application to the Beam Maser}, E. T. Jaynes and  F.W. Cummings outlined the formalism and basic mechanisms of resonant light-matter interactions \cite{JC}. In the quantum theory, the field and its quantum states can be expanded on multi-mode number states, $\phi(n_1,n_2, \ldots )$  with $n_i$ photons in the $i^\textup{th}$ field mode (using the notation of \cite{JC}). Absorption and emission processes by a single two-level system (TLS) are characterized by matrix elements of the dipole operator between the emitter's ground and excited states and of the electric field operator between field states differing in photon number in any of the field modes. The quantized electric field operator can be written as a linear combination of the annihilation and creation operators $\hat{a}_i$, $\hat{a}^\dagger_i$ for the different field modes, and hence couples $\phi(n_1,n_2, \ldots ,n_i, \ldots )$  to the states $\phi(n_1,n_2, \ldots ,n_i + 1, \ldots )$ and $\phi(n_1,n_2, \ldots ,n_i -1, \ldots )$ with strengths $\propto \sqrt{n_i+1}$ and $\sqrt{n_i}$, respectively.
 
The Jaynes-Cummings model (JCM) is the restriction of the light-matter interaction problem to the case of a two-level system coupled to a single mode of the quantized field. The JCM applies to the case of a TLS residing inside, or traveling through, an optical cavity that supports a discrete cavity mode which is nearly resonant with the TLS transition (and it typically assumes that all other cavity modes are widely detuned and hence not coupled to the TLS). The JCM has been implemented with atoms and optical cavities \cite{boca2004observation,birnbaum2005photon}, Rydberg atoms and microwave cavities \cite{filipowicz1986theory,brune1996quantum,walther2006cavity}, superconducting qubits and various microwave resonator designs \cite{fink2008climbing,hofheinz2009synthesizing}, as well as with superconducting qubits and acoustic resonators (phonons) \cite{gustafsson2014propagating,manenti2017circuit,bienfait2019phonon}.

\begin{figure}[b]
    \centering
\includegraphics[width=0.92\linewidth]{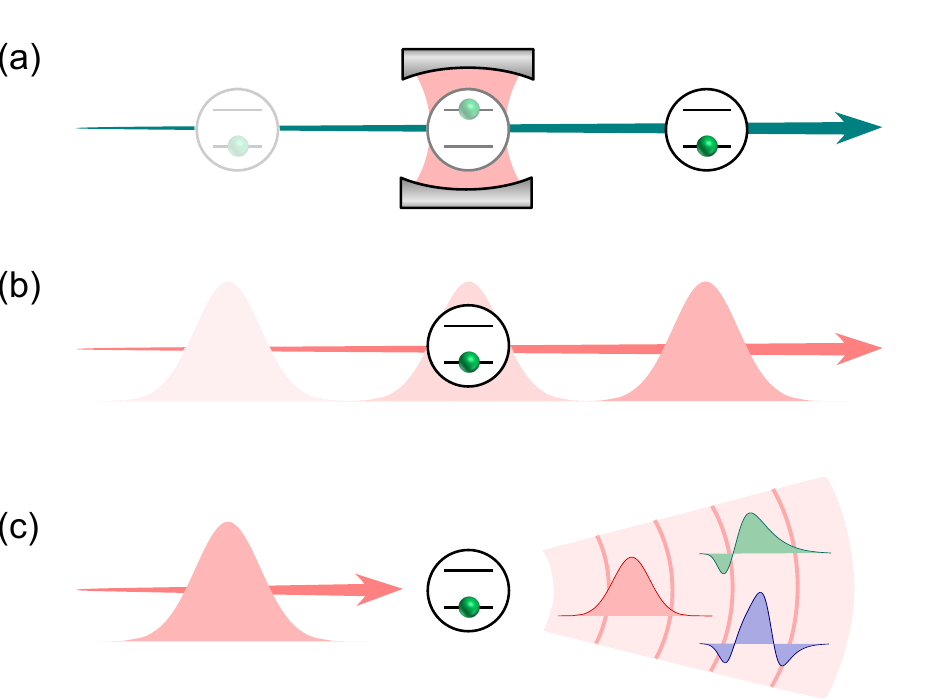}
    \caption{(a) A two-level atom passing through a single-mode cavity.(b) A single-mode pulse traveling over a two-level atom.  
    (c) A light pulse interacting with a two-level system leads to emission of photons into a continuum of field modes.}
    \label{fig:fig_JC}
\end{figure}

Using Pauli spin operators $\hat{\sigma}^{\pm}$ to represent excitation and deexcitation of the TLS, the resonant Jaynes-Cummings Hamiltonian reads
\begin{equation} \label{eq:HJCM}
H_\textup{JCM}=i\chi(\a^\dagger \hat{\sigma}^- - \a \hat{\sigma}^+).
\end{equation}
We note that $H_\textup{JCM}$ acts separately within two-dimensional subspaces $\{\ket{n,g},\ket{n-1,e}\}$ in which the dynamics periodically exchanges a single quantum between the cavity field mode and the TLS at frequency $\chi\sqrt{n}$. If different number states are populated, the quantum is exchanged at different frequencies leading to so-called collapses and revivals in the oscillating TLS excited state population \cite{rempe1987observation}.

If we want to describe the case of an atom flying through a cavity and thus traversing the standing wave field mode as depicted in Fig.~\ref{fig:fig_JC}(a), the JCM Hamiltonian Eq. \eqref{eq:HJCM} merely has to be equipped with a time-dependent value $\chi(t)$ reflecting the amplitude of the field mode at the time-dependent position of the atom.

A natural question arises: Can we employ a similar time-dependent Jaynes-Cummings Hamiltonian,
\begin{equation}\label{eq:JCMpulse}
H_\textup{JC-pulse}(t)\stackrel{?}{=}i\sqrt{\gamma}u(t)(\a^\dagger \hat{\sigma}^- - \a \hat{\sigma}^+),
\end{equation}
if a stationary TLS is coupled to an incident pulse of radiation with amplitude $\sqrt{\gamma}u(t)$, cf., Fig.~\ref{fig:fig_JC}(b)? Certainly, photons may be absorbed and emitted by the TLS like in the case of a cavity, but the propagation of the wave packet through space explores a continuum of wave number and frequency components, which are close to resonance and hence also available for the photons emitted by the TLS. A single photon may thus be absorbed from the pulse and be subsequently re-emitted into wave packets of different temporal shapes, as depicted in Fig.~\ref{fig:fig_JC}(c), and, due to the saturation nonlinearity of the TLS, scattering of a multiphoton input state may feed many output modes with different number state contents. Due to its assumption of a strictly single-mode field, the successes of the JCM within the fields of cavity QED and circuit QED seem not to extend to scattering of traveling pulses on a TLS, i.e., to the field of waveguide QED. In this article, we show that despite the apparent differences, we can apply a description similar to the JCM for a traveling pulse of radiation incident on a TLS. %\textcolor{teal}{KLAUS: I have removed Victor's  reference to solving an equation like Eq.(2). Eq.(2) is a   hypothesis, worth asking about, but it is unjustified and incorrect. Further down in the text I have also removed almost all reference to solving Eq.(2), as I prefer to give the more precise message that, rather than justifying Eq.(2), we reestablish the single mode interaction Eq.(1), but with an exact meaning (as we introduce the $u$ cavity) and with exact corrections. I am not against anybody reading the text as if Eq.(2) was almost right after all, but I just do not want to write it like that myself.}

The article is organized as follows. In Sec. II we review efforts to describe quantized light-matter interactions from a quantum field perspective. For systems sharing only a single excitation, it is possible to solve the Schr\"odinger equation exactly, and for multi-photon scattering on linear optical components, the Heisenberg equations of motion obey linear equations that can be solved for large numbers of modes. But the nonlinear scattering of even a few-photon state remains a computationally challenging problem.

In Sec. III, we introduce an effective description, in which the quantum state occupying a traveling wave packet pulse is released from a cavity mode. We can treat this artificial cavity mode and the TLS as a cascaded quantum system, and it obeys a master equation with a Hamiltonian of the same form as Eq. \eqref{eq:HJCM}, but with a time-dependent strength that differs significantly from the pulse shape that we try to model. The field leaking from the cavity and from the TLS interfere in the forward scattered radiation continuum and all excitation is gradually lost, but the master equation yields the correct time evolution of the TLS. 

In Sec. IV, we turn to the quantum state of the scattered light. This is made possible by restricting the analysis of the output field continuum to a single or a small number of output temporal modes. Our theoretical description at this stage comprises the input mode, the TLS, and one or several output mode oscillators, coupled by effective time-dependent JCM Hamiltonians. This analysis permits evaluation of the final quantum state of any output mode and hence of protocols for transforming quantum states of light, that are inspired by ideal single-mode performance in a cavity. It also reveals genuine, and useful, multimode capabilities, e.g., for the sorting of different number states into different modes \cite{yang2022deterministic} and for subtraction and transferal of a single photon from an input pulse to an orthogonal pulse mode \cite{2photon_splitting}.

In Sec. V, we adopt an interaction picture, where the TLS couples to time-dependent linear combinations of the input and output modes. In this interaction picture, we recover the interaction Hamiltonian proposed in Eq. \eqref{eq:JCMpulse}, supplemented by the coupling of the TLS to the other orthogonal mode combinations, and inclusion in the master equation of their joint loss to the continuum of modes. 

In Sec. VI, we present three examples of how the most characteristic behaviors of the Jaynes-Cummings model are recovered or modified in the case of the interaction with a travelling pulse: Rabi oscillations, collapses and revivals and photon subtraction.

Sec. VII summarizes the main results of the article.

\section{Quantum light-matter interaction}
A natural starting point to describe the quantum interaction between a propagating pulse and a TLS, but \emph{not} the one adopted in this work, is by introduction of a general representation of the quantum state of the radiation and its accompanying Schr\"odinger equation. 
In a one-dimensional (1D) waveguide, the continuous field operators $\hat{a}(t)$ evaluated at a fixed location satisfy the commutation relation $[\hat{a}(t),\hat{a}^\dagger(t^\prime)]=\delta(t-t^\prime)$, and a pure state of the field  can be expanded in the form
	\begin{equation}
		|\psi\rangle = \sum_n \int dt_1\cdots dt_n\psi_n(t_1,\cdots,t_n)\hat{a}^\dagger(t_1)\cdots \hat{a}^\dagger(t_n) |0\rangle,\label{eq:field}
	\end{equation}
where $\psi_n(t_1,\cdots,t_n)$ denotes the $n$-photon wave function. While the scattering of one- and two-photon states on a single emitter can be solved by standard scattering theory \cite{gheri1998photon,Shen:05,shen2007strongly1,shen2007strongly2}, the complex (many-body) character of the quantum state of the field makes the interaction of just a few more photons with a two-level scatterer prohibitively complicated to solve. Despite that several general formalisms have been established for an arbitrary number of photons \cite{rupasov1984rigorous,PhysRevA.92.033803,PhysRevX.7.041010,zhang2021control, Fischer2018scatteringintoone, shi2015}, their numerical implementation is restricted to the few-photon regime due to the complexity of the general problem, as well illustrated by the numerical procedures presented in \cite{Fischer2018scatteringintoone}.

Even in the case where the initial state of the incident field occupies only a single temporal mode $u(t)$ with creation operator $\hat{a}^\dagger_u=\int dt u(t)\hat{a}^\dagger(t)$, the exchange of quanta of excitation with the TLS may readily lead to population of correlated state of the general form Eq.~\eqref{eq:field}.     
Using input-output theory, it has been shown, however, that for a single mode input field the dynamics of the TLS, can be obtained from a hierarchy of coupled master equations  \cite{PhysRevA.86.013811}. This theory does not provide the quantum state of the output field, but it makes it possible to obtain mean values and correlation functions and to simulate the outcome of continuous measurements on the output field \cite{PhysRevA.86.043819,PhysRevA.96.023819,dkabrowska2019quantum}.

All the methods applicable to interactions with pulses seem to be very different from the solution of the Jaynes-Cummings model, but as we show in Sec.~\ref{sec:sec3}, we can cast the problem in such a way from the beginning that we can employ the cascaded-system-approach \cite{gardiner1993,Carmichael1993,combes2017slh}, and describe the evolution of the TLS by a JCM-like model where the output field is treated as loss.
 
In the single-mode JCM, we obtain both the state of the TLS and the field. If we should be able to obtain the quantum state of the emitted field in the form of Eq.~\eqref{eq:field}, we may consider a simpler characterization of the scattered field, i.e., the quantum state of photons in a single temporal mode $v(t)$, described by the annihilation operator $\hat{a}_v=\int dt v^*(t)\hat{a}(t)$. The density matrix for such a single-mode mixed state can be obtained by tracing out the modes orthogonal to $\hat{a}_v$, $\hat{\rho}_v=\mathrm{Tr}_{v_\perp}[|\psi\rangle\langle\psi|]$, and is given by the formal expression,
\begin{widetext}
\begin{equation}
\hat{\rho}_v=\sum_{m=0}^{\infty}\frac{1}{m!}\int dt_{1}dt_{2}\cdots{dt_m}\langle0|\hat{b}(t_m)\cdots
\hat{b}(t_2)\hat{b}(t_1)|\psi\rangle\langle\psi|\hat{b}^\dagger(t_1)\hat{b}^\dagger(t_2)\cdots
\hat{b}^\dagger(t_m)|0\rangle,
\label{eq:rhovreduced}
\end{equation}
\end{widetext}
where $\hat{b}(t)=\hat{a}(t)-v(t)\hat{a}_v$. The reduced quantum state $\hat{\rho}_v$ is the relevant quantity of interest for many quantum optical applications, and it permits, for example, direct experimental distillation via the quantum pulse gate \cite{eckstein2011quantum,serino2023realization}. In Sec.~\ref{sec:sec4} of this article, we show that the reduced density matrix $\hat{\rho}_v$ can be obtained directly and easily from an  extended version of the JCM and cascaded master equation, i.e., without ever computing the full output state $|\psi\rangle$.

\begin{figure}[b]
    \centering
\includegraphics[width=\linewidth]{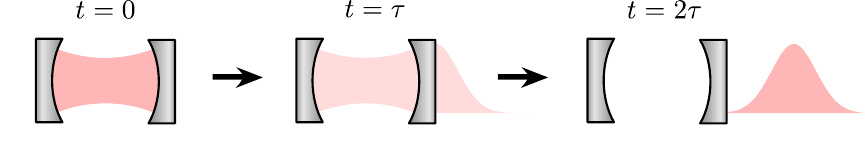}% first figure itself
    \caption{Illustration of the modeling of a traveling pulse as the emission from a single mode leaking cavity. }
    \label{fig:fig_leaking}
\end{figure}

\begin{figure*}
    \centering
\includegraphics[width=0.9\linewidth]{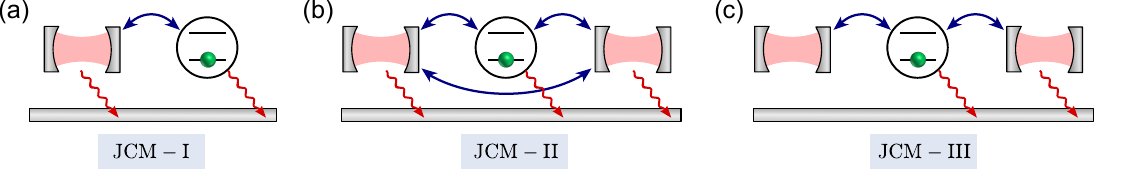}% first figure itself
    \caption{(a) Illustration of our effective model for describing the interaction between a two-level system and an input pulse, see Eq. \eqref{eq:MasterJCM-I}. (b) Illustration of our model permitting analysis of the output state in any specified temporal mode, see Eq. \eqref{eq:MasterJCM-II}. (c) When the output temporal mode (cf. panel (b)) is chosen identical to the input pulse mode, we can recover a particularly efficient description in the interaction picture, see Eq. \eqref{eq:MasterJCM-III}. The bent and curvy arrows represent the Jaynes-Cummings-like interactions given by  Hamiltonians $\H_\textup{JCM-I(II)(III)}$ and the interfering emission into the waveguide, given by the Lindblad damping operators $\L_\textup{JCM-I(II)(III)}$. The combination of the Hermitian interaction and the damping terms cause excitation to only propagate from left to right in panels (a) and (b). }\label{fig:fig_virtual}
\end{figure*}

\section{Cascaded system description of a quantum pulse incident on a two-level system}\label{sec:sec3}

We consider a pulse of radiation with a spatial amplitude $u(x,t)$ incident on the TLS. The pulse is assumed to move at a constant speed so it is equivalently specified by its spatial amplitude at a fixed early moment in time and by the time-dependent amplitude $u(t)$ at a fixed position in space. 

In Refs.~\cite{short_kiilerich,long_kiilerich, christiansen}, we argued that the interaction with the pulse is equivalent to the one obtained by positioning the TLS at a distance downstream from a cavity that emits its quantum state into a pulse of the shape $u(t)$, as depicted in Fig.~\ref{fig:fig_leaking}. For now, we assume unidirectional propagation of the field, and the constant propagation delay due to their separation $\Delta x$ can be incorporated by a formal change of time argument as if the state assigned to the cavity mode at time $t$ in reality applies at the earlier time $t-\Delta x/c$. The resulting equation treats the TLS as if it is located in the immediate vicinity outside the cavity, as shown in Fig.~\ref{fig:fig_virtual}(a). 

To leak a pulse with temporal shape $u(t)$, the cavity in Fig.~\ref{fig:fig_leaking} must have the coherent out-coupling amplitude  
\begin{align}\label{eq:g1}
g_u(t)  = \frac{u^*(t)}{\sqrt{1-\int_0^t dt'\, |u(t')|^2}}.
\end{align}
According to the theory of cascaded quantum systems 
\cite{gardiner,Carmichael1993}, this amplitude governs the Jaynes-Cummings-like interaction Hamiltonian between the cavity emitter mode and the TLS,   
\begin{equation}\label{eq:JCMcav}
H_\textup{JCM-I}(t)=\frac{i}{2}(\sqrt{\gamma}g_u(t)\a^\dagger \hat{\sigma}^- - \sqrt{\gamma}g^*_u(t) \a \hat{\sigma}^+),
\end{equation}
while the ultimate leakage of all excitation from the two systems into the waveguide is described by master equation damping terms. 

Downstream from the TLS, the radiation field is composed of the vacuum input to the waveguide, $\hat{b}_\textup{in,vac}(t)$, the field amplitude, $g_u(t)\hat{a}$, leaked by the cavity and the amplitude, $\sqrt{\gamma}\hat{\sigma}^{-}$, leaked by the TLS. These terms interfere, and downstream detection of a photon in the waveguide is accompanied by the action of the collective annihilation operator,
\begin{align}\label{eq:bout}
\hat{b}_\textup{out}(t) = \hat{b}_\textup{in,vac}(t) + g_u^*(t)\a_u +\sqrt{\gamma}\hat{\sigma}^-,
\end{align}
on the joint state of the input vacuum modes, the cavity mode, and the TLS. 

The operation of $\hat{b}_\textup{in,vac}(t)$ on vacuum is trivial, and the radiation loss represented by $\hat{b}_\textup{out}(t)$ corresponds to the action of the jump operator $\L_\textup{JCM-I}= g_u^*(t)\a_u +\sqrt{\gamma}\hat{\sigma}^{-}$ in the Lindblad master equation for the joint TLS and single cavity mode density matrix,
\begin{align}  \label{eq:MasterJCM-I}
\frac{d\rho}{dt} = -i [H_\textup{JCM-I},\rho ]+D[\L_\textup{JCM-I}]\rho,
\end{align}
where $D[\L]\rho \equiv  -\frac{1}{2}(\L^{\dagger}\L\rho + \rho\L^{\dagger}\L) + \L \rho \L^{\dagger}$. 

Eq.~\eqref{eq:MasterJCM-I} resembles a damped version of the Jaynes-Cummings model, but our system differs from the conventional Jaynes-Cummings dynamics in two important ways: (i) The coupling term has a very different time dependence than suggested in Eq. \eqref{eq:JCMpulse}, and (ii) after the passage of the pulse, both the (upstream) oscillator mode and the TLS have lost all excitations into downstream propagation of radiation away from the cavity and TLS.  

As a further interesting deviation from the usual JCM dynamics, we note that terms in the master equation coming from the unitary commutator part and from the Lindblad anticommutator terms $\frac{1}{2}(\L^{\dagger}\L\rho + \rho\L^{\dagger}\L)$ conspire so that only $\a \hat{\sigma}^+$ $(\a^\dagger \hat{\sigma}^-)$ acts on $\rho$ from the left (right). This implies the cascaded nature of the problem: no excitation propagates upstream from the TLS to the oscillator mode.

\section{The quantum state of the scattered light}
\label{sec:sec4}
Since the downstream light field is treated as loss, our master equation does not reveal its final quantum state in the same way as propagation by the Hamiltonian Eq. \eqref{eq:HJCM} explicitly provides the state of a cavity field mode during and after the Jaynes-Cummings interaction with a TLS.

The downstream radiation may be multimode and hence it may be prohibitively complicated to obtain its exact quantum state in the form of Eq.~\eqref{eq:field}. There is, however, no obstacle to the calculation of the quantum state contents of any specified single wavepacket mode $v(t)$ in the output field. To that end, we can apply a similar cavity trick as above and assume a downstream cavity with a coupling amplitude $g_v(t)$ to the waveguided field, that would fully capture the field populating the wavepacket mode $v(t)$, i.e., the reverse of the process that we employed for emission of $u(t)$.
Such a cavity should couple to the waveguide by the coupling strength 
\begin{align}\label{eq:g2}
g_v(t) = -\frac{v^*(t)}{\sqrt{\int_0^t dt'\, |v(t')|^2}}.
\end{align}

The cascaded system of the emitter cavity, the TLS, and the final pick-up cavity is depicted in Fig. \ref{fig:fig_virtual}(b). Its unidirectional dynamics is described by a Hamiltonian that exchanges quanta between all three components with their respective coupling strengths to the waveguide,
\begin{align}\label{eq:HJCM-input-output-pulse}
\begin{split}
\H_\textup{JCM-II}(t) = \frac{i}{2}&\left(\sqrt{\gamma}g_u(t)\a_u^\dagger\hat{\sigma}^-
+ \sqrt{\gamma} g_v^*(t)\hat{\sigma}^+\a_v \right. \\
 &\left. + g_u(t) g_v^*(t)\a_u^\dagger\a_v - \mathrm{H.c.} \right).
\end{split}
\end{align}
The field components that are not retrieved by the pick-up cavity, are  
governed by the annihilation operator 
\begin{align}\label{eq:bout}
\b_\textup{out}(t) = \b_\textup{in,vac}(t) + g_u^*(t)\a_u +\sqrt{\gamma}\hat{\sigma}^-+g_v^*(t)\a_v.
\end{align}
This residual loss of radiation is represented by the Lindblad damping term, 
\begin{align}\label{eq:L}
\L_\textup{JCM-II}(t) = \sqrt{\gamma}\hat{\sigma}^-+g_u^*(t)\a_u+g_v^*(t)\a_v,
\end{align}
in the master equation for the joint state of our quantum system which now consists of the TLS and two cavity modes representing the input pulse and a selected output pulse, 
\begin{align} \label{eq:MasterJCM-II}
\frac{d\rho}{dt} = -i[\H_\textup{JCM-II}(t),\rho ]+D[\L_\textup{JCM-II}(t)]\rho.
\end{align}
The scattering of a quantum pulse is calculated by assuming any initial quantum state of the input cavity-mode $(u)$, and the ground and vacuum state of the TLS and pick-up cavity mode. After solving the master equation (11), the TLS and input cavity are in the ground state and the reduced density matrix of the pick-up cavity mode represents the quantum state of the outgoing wave packet $v(t)$, as in Eq. \eqref{eq:rhovreduced}. 
 
\section{Interaction picture with respect to the free propagation of a single mode pulse}\label{sec:sec5}

We have established a Hamiltonian and a master equation that describe the interaction of an incoming pulse of quantum radiation with a quantum system, as well as the state of the field retrieved in any single mode of the output field Eq. \eqref{eq:HJCM-input-output-pulse}. This description deals with two field modes rather than the single mode in the Jaynes-Cummings model. In this section, we will show that it is possible to recover a description where the hypothesized Jaynes-Cummings Hamiltonian Eq. \eqref{eq:JCMpulse} is the dominant interaction term, but where extra components take care of the multimode nature of the problem. 

To achieve this simple description, we perform a transformation to the interaction picture defined by the cavity-cavity coupling part of the Hamiltonian
\begin{equation}
    \hat{H}_0(t) = \frac{i}{2}[g_u(t)g_v^\ast(t)\hat{a}_u^\dagger\hat{a}_v - g_v(t)g_u^\ast(t)\hat{a}_v^\dagger\hat{a}_u].
\end{equation}
At this point, we assume that the input and output virtual cavities address the same temporal shape $u(t) = v(t)$. The general case is described in \cite{christiansen}. The equations of motion for the operators in the interaction picture are
\begin{align}
    \Dot{\hat{a}}_{u,\textup{IP}}(t) &= \frac{1}{2}g_u(t)g_v^*(t) \hat{a}_{v,\textup{IP}}(t) \\
    \Dot{\hat{a}}_{v,\textup{IP}}(t) &= -\frac{1}{2}g_u^*(t) g_v(t) \hat{a}_{u,\textup{IP}}(t),
\end{align}
where $\textup{IP}$ denotes that the operator is described in the interaction picture. The solution to this coupled set of equations is
\begin{align}
    \hat{a}_{u,\textup{IP}}(t) &= \cos{\theta(t)} \hat{a}_u(0) - \sin{\theta(t)} \hat{a}_v(0) \\
    \hat{a}_{v,\textup{IP}}(t) &= \cos{\theta(t)} \hat{a}_v(0) + \sin{\theta(t)} \hat{a}_u(0),
\end{align}
where $\theta(t)$ is determined by the condition
\begin{align}
    \frac{d}{dt}\theta(t) = -\frac{1}{2}g_u(t)g_v^*(t).
\end{align}
Our assumption that $v(t) = u(t)$ guarantees that $\theta(t)$ is real, and we can show that
\begin{align} \label{eq:sin2}
    \sin^2{\theta(t)} = \int_0^t dt' |u(t')|^2,
\end{align}
by differentiating both sides of the equation with respect to time. This yields
\begin{equation}
2 \sin{\theta(t)}\cos{\theta(t)} \frac{d\theta(t)}{dt}
= |u(t)|^2,
\end{equation}
and we note that due to \eqref{eq:sin2}, the factors $\cos\theta(t)$ and $\sin\theta(t)$ are nothing but the denominators in Eqs. \eqref{eq:g1} and \eqref{eq:g2}. 

It is evident from the definition of $\theta(t)$ that $\sin^2 \theta(t=0) = 0$ and $\sin^2\theta(t\rightarrow \infty) = 1$, such that during the interaction with the pulse, the cavity operators are completely exchanged. In the interaction picture the $u$-cavity operator ``follows the pulse'', and in the case of no intermediate emitter, it represents the constant content of the pulse at all times. When the scatterer is present, the transformed Hamiltonian in the interaction picture becomes
\begin{align} \label{eq:JCM-interaction-picture}
\begin{split}
    &\hat{H}_\textup{JCM-III}(t) = i\sqrt{\gamma}u(t)(\hat{a}_{u,\textup{IP}}(t)^\dagger\hat{\sigma}^- - \hat{\sigma}^+\hat{a}_{u,\textup{IP}}(t)) \\ 
    &+ \frac{i}{2}\sqrt{\gamma}u(t)(\cot\theta-\tan\theta)(\hat{a}_{v,\textup{IP}}(t)^\dagger\hat{\sigma}^- - \hat{\sigma}^+\hat{a}_{v,\textup{IP}}(t)),
\end{split}
\end{align}
which contains in the first line precisely the naive guess, hypothesized in Eq. \eqref{eq:JCMpulse} in the Introduction. Our analysis, however, also yields the interaction with a second mode, namely the initially empty downstream cavity mode, which gets transiently occupied and hence forms a non-Markovian element in the TLS dynamics in the interaction picture, see Fig. \ref{fig:fig_virtual}(c). The pulse mode suffers no direct loss, while excitation transferred to the TLS and subsequently to the second mode leaks as revealed by evaluating the Lindblad operator in the interaction picture  
\begin{align} \label{eq:Lindblad-interaction-picture}
    \hat{L}_\textup{JCM-III}(t) = \sqrt{\gamma}\hat{\sigma}^- - (\tan\theta + \cot\theta)u(t)\hat{a}_{v,\textup{IP}}(t).
\end{align}

To summarize, we have found that the interaction between a traveling pulse and a quantum emitter has a Jaynes-Cummings-like component, which is supplemented with loss and coupling to a secondary, lossy mode in the interaction picture master equation,
\begin{align}\label{eq:MasterJCM-III}
\frac{d\rho}{dt} = -i[\H_\textup{JCM-III}(t),\rho ]+D[\L_\textup{JCM-III}(t)]\rho.
\end{align}

\begin{figure*}[t]
    \centering    \includegraphics{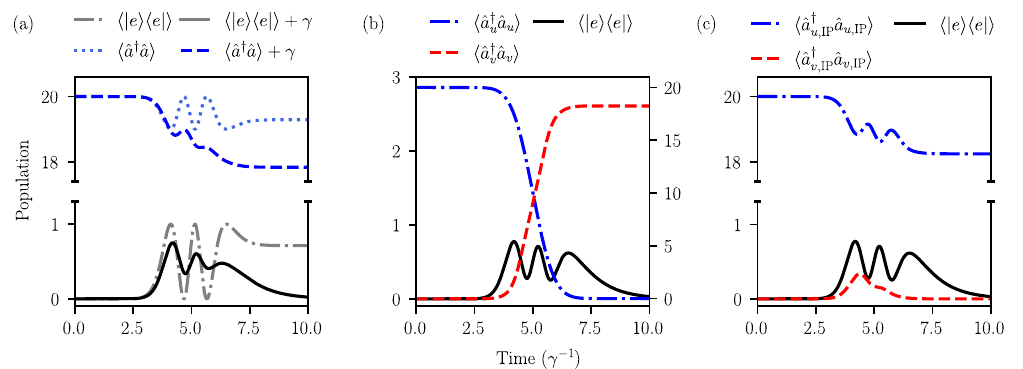}
    \caption{
    Rabi oscillations of a TLS interacting with an $|n=20\rangle$ Fock state. Panel (a) shows the TLS excitation dynamics (lower curves) and the mean number of oscillator quanta (upper curves) for the unitary JCM model with Gaussian time-dependent coupling $\chi(t)=\sqrt{\gamma}u(t)$, with $\tau = \gamma^{-1}$, and for the model with a TLS with decay rate $\gamma$. Panel (b)  shows the results of Eqs. \eqref{eq:MasterJCM-I} and \eqref{eq:MasterJCM-II} for a travelling pulse with the same Gaussian form of $u(t)$ as in panel (a). The TLS excitation dynamics is similar but differ from that in panel (a), while the excitation leaks from the upstream $u$-cavity, and is partly recovered in the downstream  $v$-cavity. Note the left y-axis pertains to the TLS and the right y-axis pertains to the cavity excitation dynamics. Panel (c) shows the same results but for the interaction picture modes applied in Eq. \eqref{eq:MasterJCM-III}. The $u_{\textup{IP}}$-mode represents the traveling pulse and its excitation dynamics is similar to the one of the oscillator mode in the damped JCM, shown in panel (a). The ancillary $v_\textup{IP}$-mode is weakly excited and causes the small quantitative discrepancies with the results in panel (a). }
\label{fig:rabioscillations}
\end{figure*}
\section{Results}

In the previous sections, we have reviewed an efficient description of the interaction between a TLS and a traveling pulse mode of quantum radiation.  We have shown that despite the multimode character of propagating fields, it is possible to recast the problem with a Jaynes-Cummings-like interaction Hamiltonian and Lindblad loss terms acting on only the TLS and one or two oscillator modes.

Our analysis has identified three different master equations \eqref{eq:MasterJCM-I}, \eqref{eq:MasterJCM-II} and \eqref{eq:MasterJCM-III}. In this section, we solve these equations, and we show how the interaction between a TLS and a traveling pulse has similarities with and deviates from the time-dependent interaction with a resonant cavity mode, with and without a dissipating TLS. 

Numerical results in this section are obtained by use of the \verb|QuTiP| toolbox \cite{qutip1, qutip2}.

\subsection{Rabi oscillations with a number state pulse}

In the master equation Eq. \eqref{eq:MasterJCM-I}, the coupling strength $g_u(t)$ differs from the shape $u(t)$ of the pulse. Note, however, that $H_\textup{JCM-I}$ acts on a quantum state with progressively fewer photons, governed by the same factor as the denominator in Eq. \eqref{eq:g1}, and hence its strength follows the value of $u(t)$ in an average sense. 

Figure \ref{fig:rabioscillations}(a) shows the photon number and TLS excitation probability according to the unitary JCM dynamics of Eq. \eqref{eq:HJCM}  with a time-dependent coupling, $\chi(t) = \sqrt{\gamma}u(t)$, with a Gaussian temporal width $\tau = \gamma^{-1}$, for the cavity mode initially in an $|n=20\rangle$ Fock state. The figure also shows the solution of the master equation incorporating decay from the excited state of the TLS with rate $\gamma$, where we note a loss of a little more than $2$ photons. 

Figure \ref{fig:rabioscillations}(b) shows the results for an incident Gaussian wave packet, $u(t)$, with an $|n=20\rangle$ Fock state of radiation scattering on the TLS, as in Eq. \eqref{eq:MasterJCM-I}. The TLS excited state population yields similar damped Rabi oscillations as in panel (a), accompanying the rapid release of the field from the $u$-cavity. Fig. \ref{fig:rabioscillations}(b) also shows the population emerging in the outgoing wavepacket $v(t)$ [chosen here on the same form as $u(t)$] after the interaction with the TLS, as in Eq. \eqref{eq:MasterJCM-II}. We note a loss of a little less than $2$ quanta of radiation to other modes.

Figure \ref{fig:rabioscillations}(c) shows the same TLS excitation dynamics but calculated according to Eq. \eqref{eq:MasterJCM-III}. The interaction picture annihilation operator $\hat{a}_{u,\mathrm{IP}}(t)$ represents the $u(t)$ wave packet mode both before and after the interaction with the TLS, and its population dynamics (upper dot dashed curve) resembles the conventional JCM dynamics in presence of TLS decay, shown in panel (a). The interaction picture annihilation operator $\hat{a}_{v,\mathrm{IP}}(t)$ represents an orthogonal mode that is weakly excited and thus modifies the TLS dynamics during the interaction. 

Although, we obtain results showing similar features with the single-mode time-dependent Jaynes-Cummings model with damping and with our modified JCM-like master equations, there are quantitative discrepancies. These are due to the multimode character of the non-linear scattering of traveling pulse, which is only taken fully into account in the extended versions of the JCM model presented here.  

\subsection{Number state vs. coherent state dynamics. Collapses and revivals. }

The cascaded nature of the master equation, Eqs. \eqref{eq:MasterJCM-I} and \eqref{eq:MasterJCM-II}, ensures that the combination of the commutator and the Lindblad anti-commutator terms are responsible for excitations propagating only downstream (as only annihilation and no creation operators for the upstream cavity act on $\rho$ from the left, and vice versa from the right). This has another striking consequence: if the cavity field mode is initially in a coherent state with amplitude $\alpha_0$, it remains a (damped) coherent state $|\alpha(t)\rangle$ for all later times. The TLS thus evolves as if a classical pulse drives it with a time-dependent field amplitude $\alpha(t) g_u^*(t) =\alpha_0 u(t)$ \cite{PhysRevA.86.043819,long_kiilerich}. In addition, the TLS decays to the waveguide with the rate $\gamma$, and the system may thus show damped Rabi oscillations but no trace of the collapse and revival phenomenon, which is a crucial feature of the interaction with a coherent state in a cavity mode. 

Conversely, for an initial number state, which would exhibit perfect Rabi oscillations in the JCM, our Eq. \eqref{eq:MasterJCM-I} evolves the system through a gradual loss of the initial excitation. It hence explores a distribution of excitation subspaces with a corresponding distribution of Rabi oscillation frequencies $g\sqrt{n}$. This could potentially lead to prediction and observation of collapse and revival occurring in the interaction with a traveling pulse in a number state, which would be the complete reverse of roles of coherent states and number states in the JCM. 

The gradual loss of excitation and broadening of the number state distribution in the $u$-mode, does not, however, occur to the same extent, when we solve the dynamics in the interaction picture, cf., the upper curve in Fig.~\ref{fig:rabioscillations}(c). Since our Eqs. \eqref{eq:MasterJCM-I} and \eqref{eq:MasterJCM-III} yield identical TLS dynamics, despite the structure of Eq.~\eqref{eq:MasterJCM-I}, we should hence not expect to observe the Rabi oscillation collapses and revivals, while we should expect to observe damped oscillations due to the decay of the TLS.      

\subsection{Further dissipation, non-chiral coupling}

In the previous paragraphs, we have mentioned the possibility of incorporating decay and dissipation in the JCM through damping terms in a master equation. Similarly, it is straightforward to supplement our master equations with further damping terms of the same form $D[\L_i]\rho$ as the Lindblad terms already present in Eqs. \eqref{eq:MasterJCM-I}, \eqref{eq:MasterJCM-II} and \eqref{eq:MasterJCM-III}. 

This is not the place for an exhaustive review of such damping terms, but let us consider here what may have seemed a severe restriction of our analyses - a perfect unidirectional interaction between the TLS and the propagating fields. Most emitters couple to fields propagating in both directions along waveguides, and if we want to incorporate the possibility of reflection from the emitter, it is only a matter of adding a corresponding Lindblad term $D[\L_\mathrm{refl}]\rho$ with $\L_\mathrm{refl} = \sqrt{\gamma'}\hat{\sigma}^-$ to our master equations. The analysis of the TLS dynamics and the analysis of the forward propagating field proceed by solving the master equations with this extra term. 

\begin{figure}
    \centering
    \includegraphics{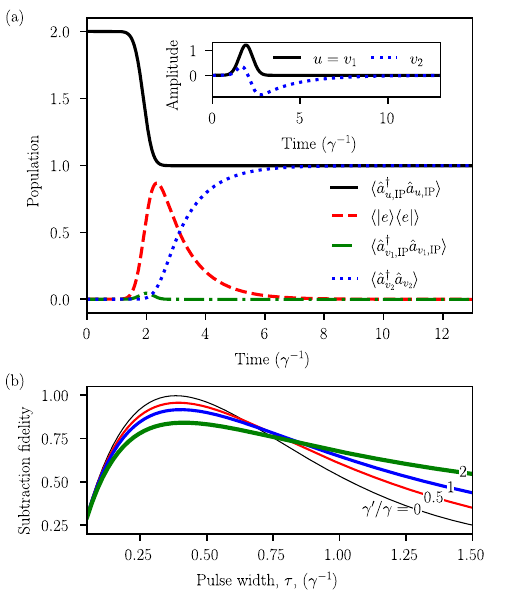}
    \caption{Panel (a) presents the solution of Eq. \eqref{eq:MasterJCM-III} for the interaction of a TLS with a Gaussian pulse of width $\tau =0.3799\gamma^{-1}$ and two quanta of excitation. The calculations show the removal of one quantum  from the input mode and the emergence of one quantum in a new, orthogonal temporal mode (these modes are shown in the inset). The density matrix elements for the modes (not shown) reveal that the two output modes are, indeed, populating $|n=1\rangle$ Fock states. Panel (b)  shows the fidelity of the subtraction of a single quantum from the incident pulse as function of the pulse duration, and for different values of the TLS emission rate in the direction opposite to the pulse propagation. Perfect subtraction only occurs in the case of perfect chiral coupling of the TLS to the travelling pulse.}   
    \label{fig:2psplitting}
\end{figure}
In Fig.~\ref{fig:2psplitting}(a), we show a particular process that occurs in the case of perfect chiral coupling of a TLS to a two-photon wave packet, namely the very precise removal of exactly one photon from the incident pulse $u(t)$ (black curve) \cite{2photon_splitting}. The dynamics is obtained in the interaction picture, using Eq. \eqref{eq:MasterJCM-III}. Naively, this process is well understood from the JCM Rabi oscillation dynamics, causing excitation of the TLS with a $\pi$-pulse and subsequent emission of a single quantum of excitation. However, in our case, the TLS is constantly coupled to and emitting into the waveguide modes. As shown by the red curve in Fig. \ref{fig:2psplitting}(a), the TLS excitation is not complete, but for the right parameter values, over time it emits one quantum of excitation (blue dotted curve) into a temporal pulse mode $v_2(t)$ which is orthogonal to $u(t)$ (see the modes in the inset). This effective splitting of a number state pulse into two one-photon states may have useful implications in quantum information processing with light, but will it only work if we can ensure the unidirectional coupling of the TLS to the traveling fields? The answer is given in Fig.~\ref{fig:2psplitting}(b), where we show the fidelity of the photon subtraction process as a function of the duration of the incident Gaussian 2-photon pulses. The different curves are obtained for different values of the emission rate in the direction of reflection from the TLS. Only for the perfect chiral coupling do we find a pulse duration with perfect subtraction.

\section{Conclusion}

The textbook  Jaynes-Cummings model assumes a single quantized field mode and, hence, it does not describe the interaction between a two-level system and a traveling pulse of light due to the multi-mode character of the field continuum available for the scattered photons. We have shown that in a cascaded open-system approach it is possible to obtain master equations \eqref{eq:MasterJCM-I}, \eqref{eq:MasterJCM-II} and \eqref{eq:MasterJCM-III} of similar form and complexity as the ones describing the Jaynes-Cummings model. We remind the reader that the cavities depicted in Figs. \ref{fig:fig_leaking} and \ref{fig:fig_virtual} are only introduced to aid the theoretical description of what are propagating temporal modes in the actual experiment. It is through that effective description that we recover single- and few-mode variants of the Jaynes-Cummings interaction. With the assumption of the validity of the Born-Markov approximation of the emission of the TLS into the waveguide, our master equations are exact. 

By our discussions of the different forms of the master equations and by our presentation of results for specific processes we have demonstrated the most important consequences of the modifications to the Jaynes-Cummings model in the case of propagating light. We are delighted that by so simple modifications of the theory, the 60-year-old Jaynes-Cummings model may still be counted on and provide intuition and quantitative predictions for a wide range of future applications.

\section{Acknowledgements}
We thank Anton L. Andersen and Danil Kornovan for their helpful discussions. The authors acknowledge support from the Danish National Research Foundation through the Center of Excellence for Complex Quantum Systems and the Center of Excellence for Hybrid Quantum Networks (Grant agreements No. DNRF156 and DNRF139) and the Carlsberg Foundation through the Semper Ardens project QCooL.

\bibliography{bibliography.bib}

\end{document}